\documentclass[english]{article}
\usepackage[T1]{fontenc}
\usepackage[latin1]{inputenc}
\usepackage{amsmath}
\usepackage{setspace}
\usepackage{amssymb}

\makeatletter
\newcommand{\lyxaddress}[1]{
\par {\raggedright #1
\vspace{1.4em}
\noindent\par}
}

\usepackage{babel}
\makeatother
\begin{document}

\title{Quaternion Dirac Equation and Supersymmetry}

\author{Seema Rawat$^{\text{(1)}}$ and O. P. S. Negi$^{\text{(2)}}$}

\maketitle

\lyxaddress{\begin{center}$^{\text{(1)}}$Department of Physics\\
Govt. P.G.College\\
Ramnagar (Nainital), U.A.\par\end{center}}

\lyxaddress{\begin{center}$^{\text{(2)}}$Department of Physics\\
Kumaun University\\
S. S. J. Campus\\
Almora- 263601, U.A.\par\end{center}}

\lyxaddress{\begin{center}E-mail:- $^{\text{(1)}}$rawatseema1@rediffmail.com\\
$^{\text{(2)}}$ops\_negi@yahoo.co.in. \par\end{center}}

\begin{abstract}
Quaternion Dirac equation has been analysed and its supersymetrization
has been discussed consistently. It has been shown that the quaternion
Dirac equation automatically describes the spin structure with its
spin up and spin down components of two component quaternion Dirac
spinors associated with positive and negative energies. It has also
been shown that the supersymmetrization of quaternion Dirac equation
works well for different cases associated with zero mass, non zero
mass, scalar potential and generalized electromagnetic potentials.
Accordingly we have discussed the splitting of supersymmetrized Dirac
equation in terms of electric and magnetic fields.
\end{abstract}

\section{Introduction}

~~~Symmetries are one of the most powerful tools in theoretical
physics. The two component formulation of complex numbers, and the
noncommutative algebra of quaternions, are possibly the two most important
discoveries in mathematics. Quaternions were very first example of
hyper complex numbers having the significant impacts on mathematics
\& physics \cite{key-1} . Because of their beautiful and unique properties
quaternions attracted many to study the laws of nature over the field
of these numbers. Quaternions are already used in the context of special
relativity \cite{key-2}, electrodynamics \cite{key-3,key-4}, Maxwell's
equation \cite{key-5}, quantum mechanics \cite{key-6,key-7}, Quaternion
Oscillator \cite{key-8}, gauge theories \cite{key-9,key-10} , Supersymmetry
\cite{key-11}and many other branches of Physics \cite{key-12} and
Mathematics \cite{key-13}. On the other hand supersymmetry (SUSY)
is described as the symmetry of bosons and fermions \cite{key-14,key-15,key-16}.
Gauge Hierarchy problem, not only suggests that the SUSY exists but
put an upper limit on the masses of super partners \cite{key-17,key-18}.
The exact SUSY implies exact fermion-boson masses, which has not been
observed so far. Hence it is believed that supersymmetry is an approximate
symmetry and it must be broken \cite{key-19,key-20}. We have considered
following two motivations to study the higher dimensional supersymmetric
quantum mechanics \cite{key-21} over the field of Quaternions. 

1. Supersymmetric field theory can provide us realistic models of
particle physics which do not suffer from gauge hierarchy problem
and role of quaternions will provide us simplex and compact calculation
accordingly.

2. Quaternions super symmetric quantum mechanics can give us new window
to understand the behavior of supersymmetric partners and mechanism
of super symmetry breaking etc.

3. Quaternions are capable to deal the higher dimensional structure
and thus include the theory of monopoles and dyons \cite{key-22,key-23}.

Keeping these facts in mind and to observe the role of quaternions
in supersymmetry, we have developed the quaternion Dirac equation
parallel to Dirac Pauli representation.It has been shown that the
quaternion Dirac equation automatically describes the spin structure
with its spin up and spin down components of two component quaternion
Dirac spinors associated with positive and negative energies. We have
obtained the free particle solutions of quaternion Dirac equation
for quaternion, complex and real spinor representations. It has been
shown that one component quaternion valued Dirac spinor is isomorphic
to two component complex spinors or four components real representation.
It has also been shown that the supersymmetrization of quaternion
Dirac equation works well for different cases associated with zero
mass, non zero mass, scalar potential and generalized electromagnetic
potentials. Accordingly we have discussed the splitting of supersymmetrized
Dirac equation in terms of electric and magnetic fields. Accordingly,
the super charges are calculated in all cases and it is shown that
the Hamiltonian operator commutes with the super charges and the relations
between the Schroedinger Hamiltonians and Dirac Hamiltonians are discussed
in terms of super charges for different cases.

\section{Definition}

A quaternion $\phi$ is expressed as

\begin{eqnarray}
\phi & = & e_{\text{0}}\phi_{0}+e_{1}\phi_{1}+e_{2}\phi_{2}+e_{3}\phi_{3}\label{eq:1}\end{eqnarray}
Where $\phi_{0},\phi_{1},\phi_{2},\phi_{3}$ are the real quartets
of a quaternion and $e_{0},e_{1},e_{2},e_{3}$ are called quaternion
units and satisfies the following relations,

\begin{eqnarray}
e_{0}^{2} & = & e_{0}=1,\nonumber \\
e_{0}e_{i} & = & e_{i}e_{0}=e_{i}(i=1,2,3),\nonumber \\
e_{i}e_{j} & = & -\delta_{ij}+\varepsilon_{ijk}e_{k}(i,j,k=1,2,3),\label{eq:2}\end{eqnarray}
where $\delta_{ij}$ is the Kronecker delta and $\varepsilon_{ijk}$
is the three index Levi- Civita symbols with their usual definitions.
The quaternion conjugate $\bar{\phi}$ is then defined as \begin{eqnarray}
\bar{\phi} & = & e_{\text{0}}\phi_{0}-e_{1}\phi_{1}-e_{2}\phi_{2}-e_{3}\phi_{3}\label{eq:3}\end{eqnarray}
Here $\phi_{0}$is real part of the quaternion defined as\begin{align}
\phi_{0} & =Re\,\,\phi=\frac{1}{2}(\bar{\phi}+\phi)\label{eq:4}\end{align}
If $Re\,\,\phi=\phi_{0}=0$ , then $\phi=-\bar{\phi}$ and imaginary
$\phi$ is called pure quaternion and is written as

\begin{eqnarray}
Im\,\,\phi & = & e_{1}\phi_{1}+e_{2}\phi_{2}+e_{3}\phi_{3}\label{eq:5}\end{eqnarray}
The norm of a quaternion is expressed as 

\begin{eqnarray}
N(\phi) & = & \bar{\phi}\phi=\phi\bar{\phi}=\phi_{0}^{2}+\phi_{1}^{2}+\phi_{2}^{2}+\phi_{3}^{2}\geq0\label{eq:6}\end{eqnarray}
and the inverse of a quaternion is described as

\begin{eqnarray}
\phi^{-1} & = & \frac{\bar{\phi}}{\left|\phi\right|}\label{eq:7}\end{eqnarray}
While the quaternion conjugation satisfies the following property

\begin{eqnarray}
\overline{(\phi_{1}\phi_{2})} & = & \bar{\phi_{2}}\bar{\phi}_{1}\label{eq:8}\end{eqnarray}
The norm of the quaternion (\ref{eq:6}) is positive definite and
enjoys the composition law

\begin{eqnarray}
N(\phi_{1}\phi_{2}) & = & N(\phi_{1})N(\phi_{2})\label{eq:9}\end{eqnarray}
Quaternion (\ref{eq:1}) is also written as $\phi=(\phi_{0},\overrightarrow{\phi})$
where $\overrightarrow{\phi}=$$e_{1}$$\phi_{1}+e_{2}\phi_{2}+e_{3}$$\phi_{3}$
is its vector part and $\phi_{0}$is its scalar part. The sum and
product of two quaternions are

\begin{eqnarray}
(\alpha_{0},\overrightarrow{\alpha})+(\beta_{0},\overrightarrow{\beta}) & = & (\alpha_{0}+\beta_{0},\overrightarrow{\alpha}+\overrightarrow{\beta})\nonumber \\
(\alpha_{0},\overrightarrow{\alpha})\,(\beta_{0},\overrightarrow{\beta}) & = & (\alpha_{0}\beta_{0}-\overrightarrow{\alpha}.\overrightarrow{\beta},\,\alpha_{0}\overrightarrow{\beta}+\beta_{0}\overrightarrow{\alpha}+\overrightarrow{\alpha}\times\overrightarrow{\beta})\label{eq:10}\end{eqnarray}
Quaternion elements are non-Abelian in nature and thus represent a
division ring.

\section{Quaternion Dirac Equation}

Let us define respectively the space-time four vector, momentum four
vector and four differential operator as quaternion in the following
manner on using natural units $c=\hbar=1$ and $i=\sqrt{-1}$through
out the text;\begin{eqnarray}
x & = & e_{1}x_{1}+e_{2}x_{2}+e_{3}x_{3}+x_{4}=-i\, t+e_{1}x_{1}+e_{2}x_{2}+e_{3}x_{3},\label{eq:11}\\
p & = & e_{1}p_{1}+e_{2}p_{2}+e_{3}p_{3}+p_{4}=-i\, E+e_{1}p_{1}+e_{2}p_{2}+e_{3}p_{3},\label{eq:12}\\
\boxdot & = & e_{1}\partial_{1}+e_{2}\partial_{2}+e_{3}\partial_{3}+\partial_{4}=-i\,\frac{\partial}{\partial t}+e_{1}\frac{\partial}{\partial x_{1}}+e_{2}\frac{\partial}{\partial x_{2}}+e_{3}\frac{\partial}{\partial x_{3}}.\label{eq:13}\end{eqnarray}
We may define accordingly the quaternion conjugate of these physical
quantities as follows;

\begin{eqnarray}
\overline{x} & = & -e_{1}x_{1}-e_{2}x_{2}-e_{3}x_{3}+x_{4}=-i\, t-e_{1}x_{1}-e_{2}x_{2}-e_{3}x_{3},\label{eq:14}\\
\overline{p} & = & -e_{1}p_{1}-e_{2}p_{2}-e_{3}p_{3}+p_{4}=-i\, E-e_{1}p_{1}-e_{2}p_{2}-e_{3}p_{3},\label{eq:15}\\
\overline{\boxdot} & = & -e_{1}\partial_{1}-e_{2}\partial_{2}-e_{3}\partial_{3}+\partial_{4}=-i\,\frac{\partial}{\partial t}-e_{1}\frac{\partial}{\partial x_{1}}-e_{2}\frac{\partial}{\partial x_{2}}-e_{3}\frac{\partial}{\partial x_{3}}.\label{eq:16}\end{eqnarray}
Using the quaternion multiplication rule (\ref{eq:2}) we may write
the norm of the above quaternion valued four vectors as 

\begin{eqnarray}
x\overline{x}=\overline{x}x & = & x_{1}^{2}+x_{2}^{2}+x_{3}^{2}+x_{4}^{2}=x_{1}^{2}+x_{2}^{2}+x_{3}^{2}-t^{2}\label{eq:17}\\
p\overline{p}=\overline{p}p & = & p_{1}^{2}+p_{2}^{2}+p_{3}^{2}+p_{4}^{2}=p_{1}^{2}+p_{2}^{2}+p_{3}^{2}-E^{2}\label{eq:18}\\
\boxdot\overline{\boxdot}=\overline{\boxdot}\boxdot & = & \partial_{1}^{2}+\partial_{2}^{2}+\partial_{3}^{2}+\partial_{4}^{2}=\frac{\partial^{2}}{\partial x_{1}^{2}}+\frac{\partial^{2}}{\partial x_{2}^{2}}+\frac{\partial^{2}}{\partial x_{3}^{2}}-\frac{\partial^{2}}{\partial t^{2}}.\label{eq:19}\end{eqnarray}
Equation ( \ref{eq:19}) can also be related with the D'Alembertian
operator in the fallowing manner i.e.

\begin{eqnarray}
\square & = & \boxdot\overline{\boxdot}=\overline{\boxdot}\boxdot=\frac{\partial^{2}}{\partial x_{1}^{2}}+\frac{\partial^{2}}{\partial x_{2}^{2}}+\frac{\partial^{2}}{\partial x_{3}^{2}}-\frac{\partial^{2}}{\partial t^{2}}=-\frac{\partial^{2}}{\partial t^{2}}+\triangledown^{2}.\label{eq:20}\end{eqnarray}
As such we may write the quaternion form of the Klein Gordon equation
as follows,

\begin{eqnarray}
(\square-m^{2})\phi_{\alpha} & = & (\boxdot\overline{\boxdot}-m^{2})\phi_{\alpha}=(\overline{\boxdot}\boxdot-m^{2})\phi_{\alpha}=0\,(\alpha=0,1,2,3)\label{eq:21}\end{eqnarray}
where $\phi_{\alpha}$are the components of quaternion scalar field
( \ref{eq:1}). Now we may describe the quaternion form of Dirac equation
from the following Schrodinger equation

\begin{eqnarray}
\widehat{H_{D}}\psi & = & i\,\frac{\partial\psi}{\partial t}\label{eq:22}\end{eqnarray}
where 

\begin{eqnarray}
\widehat{H_{D}} & = & \sum_{l=1}^{3}\alpha_{l}p_{l}+\beta m\label{eq:23}\end{eqnarray}
and $\alpha$ and $\beta$ are the arbitrary coefficients and thus
satisfy the following properties in order to obtain the Klein Gordon
equation (\ref{eq:21}) i.e.\begin{eqnarray}
\alpha_{l}^{2} & = & 1;\,\,\,\,\,\beta^{2}=1(l=1,2,3);\nonumber \\
\alpha_{l}\beta+\beta\alpha_{l} & = & 0;\,\,\,\,\,\,\,\alpha_{l}\alpha_{m}+\alpha_{m}\alpha_{l}=0\label{eq:24}\end{eqnarray}
along with the following properties to maintain the Hermiticity of
Dirac Hamiltonian given by equation (\ref{eq:22}) i.e.;

\begin{eqnarray}
H_{D}^{\dagger} & = & H_{D}\Rightarrow\alpha_{l}^{\dagger}=\alpha_{l};\,\beta^{\dagger}=\beta\label{eq:25}\end{eqnarray}
Thus, we may define $2\times2$ quaternion valued $\alpha$ and $\beta$
matrices satisfying the properties (\ref{eq:24},\ref{eq:25})as follows
\cite{key-24} i.e.

\begin{eqnarray}
\alpha_{l}=\left[\begin{array}{cc}
0 & ie_{l}\\
ie_{l} & 0\end{array}\right];\,\,\,\,\,\,\,\, & \beta= & \left[\begin{array}{cc}
1 & 0\\
0 & -1\end{array}\right]\label{eq:26}\end{eqnarray}
where the quaternion basis elements $e_{l}$ satisfy the properties
given by equation (\ref{eq:2}). We may now introduce the plain wave
solution of Dirac equation 

\begin{eqnarray}
(\sum_{l=1}^{3}\alpha_{l}p_{l}+\beta m)\psi & = & E\,\psi\label{eq:27}\end{eqnarray}
as

\begin{eqnarray}
\psi\left(x,t\right) & = & \psi e^{ip_{\mu}x^{\mu}}\label{eq:28}\end{eqnarray}
where $\psi$ is the quaternion valued Dirac spinor given by

\begin{eqnarray}
\psi & = & \psi_{0}+e_{1}\psi_{1}+e_{2}\psi_{2}+e_{3}\psi_{3}\label{eq:29}\end{eqnarray}
and may be decomposed as $\psi=\left[\begin{array}{c}
\psi_{a}\\
\psi_{b}\end{array}\right]=\left[\begin{array}{c}
\psi_{0}\\
\psi_{1}\\
\psi_{2}\\
-\psi_{3}\end{array}\right]$ as the two or four components Dirac spinor associated with the sympletic
representation of quaternion as $\psi=\psi_{a}+e_{2}\psi_{b}$in terms
of complex and accordingly $\psi_{a}=\psi_{0}+e_{1}\psi_{1}$ and
$\psi_{b}=\psi_{2}-e_{1}\psi_{3}$over the field of real number representations.
In other words we can write one component quaternion valued Dirac
spinor which is isomorphic to two component complex spinors or four
components real representation. In describing the theory of Dirac
equation it is customary to take the Dirac spinor as the four component
spinors with complex coefficients. Hence in equation (\ref{eq:29})
the spinor $\psi$ may be described as bi-quaternion valued where
all its components i.e $\psi_{0},\psi_{1},\psi_{2},\psi_{3}$ are
complex ones and the complex quantity $i=\sqrt{-1}$ commutes with
all the quaternion basis elements $e_{0}=1,e_{1},e_{2},e_{3}$.Using
equations (\ref{eq:26}-\ref{eq:29}) we get the following two equations
as

\begin{eqnarray}
E\psi_{a} & = & m\psi_{a}+ie_{l}.p_{l}\psi_{b};\nonumber \\
E\psi_{b} & = & -m\psi_{a}+ie_{l}.p_{l}\psi_{b}\label{eq:30}\end{eqnarray}

or\begin{eqnarray}
\left(E-m\right)\psi_{a} & =ie_{l}.p_{l} & \psi_{b};\nonumber \\
\left(E+m\right)\psi_{b} & = & ie_{l}.p_{l}\psi_{a}.\label{eq:31}\end{eqnarray}
As such we may obtain the following types of four spinor amplitudes
of Dirac spinors i.e.-

(i) One component spinor amplitudes-

\begin{eqnarray}
\psi^{1}= & (1+e_{2}.\frac{ie_{l}\, p_{l}}{E+m}) & \,(Energy=+ive,\, spin=\uparrow);\nonumber \\
\psi^{2}= & (1+e_{2}.\frac{ie_{l}\, p_{l}}{E+m})e_{1} & \,(Energy=+ive,\, spin=\downarrow);\nonumber \\
\psi^{3}= & (e_{2}-\frac{ie_{l}\, p_{l}}{E+m}) & \,(Energy=-ive,\, spin=\uparrow);\nonumber \\
\psi^{4}= & (e_{2}-\frac{ie_{l}\, p_{l}}{E+m})e_{1} & \,(Energy=-ive,\, spin=\downarrow).\label{eq:32}\end{eqnarray}
 (ii) Two component spinor amplitudes-

\begin{eqnarray}
\psi^{1} & = & \left(\begin{array}{c}
1\\
\frac{ie_{l}\, p_{l}}{E+m}\end{array}\right)\,(Energy=+ive,\, spin=\uparrow);\nonumber \\
\psi^{2} & = & \left(\begin{array}{c}
1\\
\frac{ie_{l}\, p_{l}}{E+m}\end{array}\right)e_{1}\,(Energy=+ive,\, spin=\downarrow);\nonumber \\
\psi^{3} & = & \left(\begin{array}{c}
\frac{-ie_{l}\, p_{l}}{E+m}\\
1\end{array}\right)\,(Energy=-ive,\, spin=\uparrow);\nonumber \\
\psi^{4} & = & \left(\begin{array}{c}
\frac{-ie_{l}\, p_{l}}{E+m}\\
1\end{array}\right)e_{1}\,(Energy=-ive,\, spin=\downarrow).\label{eq:33}\end{eqnarray}

(iii) Four component spinor amplitudes are obtained by restricting
the direction of propagation along $Z-axis$ i.e $p_{x}=P_{y}=0$
(direction of propagation) and on substituting $e_{l}=-i\sigma_{l}$and
$\sigma_{1}=\left(\begin{array}{cc}
0 & 1\\
1 & 0\end{array}\right),\,\,\sigma_{2}=\left(\begin{array}{cc}
0 & -i\\
i & 0\end{array}\right)\,\,\sigma_{3}=\left(\begin{array}{cc}
1 & 0\\
0 & -1\end{array}\right)$ along with the usual definitions of spin up and spin down amplitudes
of spin i.e.

\begin{eqnarray}
\psi^{1} & = & \left(\begin{array}{c}
1\\
0\\
\frac{\left|\vec{p}\right|}{E+m}\\
0\end{array}\right)(Energy=+ive,\, spin=\uparrow);\nonumber \\
\psi^{2} & = & \left(\begin{array}{c}
0\\
1\\
0\\
-\frac{\left|\vec{p}\right|}{E+m}\end{array}\right)(Energy=+ive,\, spin=\downarrow);\nonumber \\
\psi^{3} & = & \left(\begin{array}{c}
-\frac{\left|\vec{p}\right|}{E+m}\\
0\\
1\\
0\end{array}\right)(Energy=-ive,\, spin=\uparrow);\nonumber \\
\psi^{4} & = & \left(\begin{array}{c}
0\\
\frac{\left|\vec{p}\right|}{E+m}\\
0\\
1\end{array}\right)(Energy=-ive,\, spin=\downarrow).\label{eq:34}\end{eqnarray}
As such we may obtain the solution of quaternion Dirac equation in
terms of one component quaternion, two component complex and four
component real spinor amplitudes. Equation (\ref{eq:34}) is same
as obtained in the case of usual Dirac equation. Thus we may interpret
that the $N=1$ quaternion spinor amplitude is isomorphic to $N=2$
complex and $N=4$real spinor amplitude solution of Dirac equation.
We can accordingly interpret the minimum dimensional representation
for Dirac equation is $N=1$ in quaternionic case, $N=2$ in complex
case and $N=4$ for real number field.

\section{Super-symmetrization of Quaternion Dirac Equation }

~~~~The quaternion free particle Dirac equation given by equation
(\ref{eq:27}) may now be directly written in following covariant
form,

\begin{eqnarray}
i\,\,\gamma_{\mu\,\,}\partial_{\mu}\psi(x,t) & = & m\,\psi(x,t)\,\,(\mu=0,1,2,3)\label{eq:35}\end{eqnarray}
where we have defined the following representation of gamma matrices
in terms of quaternion basis elements i.e.

\begin{eqnarray}
\gamma_{0}=\beta & = & \left[\begin{array}{cc}
1 & 0\\
0 & -1\end{array}\right];\nonumber \\
\gamma_{l} & =\beta\alpha_{l}= & \left[\begin{array}{cc}
0 & ie_{l}\\
-ie_{l} & 0\end{array}\right]\,\,\,\,(l=1,2,3).\label{eq:36}\end{eqnarray}
along with the following properties associated therein i.e.

\begin{eqnarray}
\gamma_{\mu}\gamma_{\nu}+\gamma_{\nu}\gamma_{\mu} & = & -2g_{\mu\nu}\nonumber \\
g_{\mu\nu} & = & (-1,+1,+1,+1)\,\forall\mu=\nu=0,1,2,3\nonumber \\
g_{\mu\nu} & = & 0\,\,\forall\mu\neq\nu.\label{eq:37}\end{eqnarray}

Let us discuss the super-symmetrization in following three cases.

\textbf{Case I}: For mass less free particle i.e.$m=0$ and external
potential $\Phi=0$; Equation (\ref{eq:35} )becomes 

\begin{eqnarray}
i\,\,\gamma_{\mu\,\,}\partial_{\mu}\psi(x,t) & = & 0\label{eq:38}\end{eqnarray}
Let us consider the following solutions of this equation as

\begin{eqnarray}
\psi(x,t) & = & \psi(x)\, e^{i\,(\overrightarrow{p}\cdot\overrightarrow{x}-Et)}\label{eq:39}\end{eqnarray}
Thus equation(\ref{eq:35}) reduces to \begin{eqnarray}
(\gamma_{0}E\,-\gamma_{1}p_{1}-\gamma_{2}p_{2}-\gamma_{3}p_{3}) & \psi(x)= & 0\label{eq:40}\end{eqnarray}
which takes the following form

\begin{eqnarray}
\left\{ \left[\begin{array}{cc}
1 & 0\\
0 & -1\end{array}\right]E-e_{l}\left[\begin{array}{cc}
0 & i\\
-i & 0\end{array}\right]p_{l}\right\} \left\{ \begin{array}{c}
\psi_{a}\\
\psi_{b}\end{array}\right\}  & = & 0\label{eq:41}\end{eqnarray}
where $\psi_{a}=\psi_{0}+e_{1}\psi_{1}$and $\psi_{b}=\psi_{2}-e_{1}\psi_{3}$.
We thus obtain the following coupled equations

\begin{eqnarray}
\widehat{A}\psi_{a}(x) & = & E\,\psi_{b}(x)\nonumber \\
\widehat{A}^{\dagger}\psi_{b}(x) & = & E\,\psi_{a}(x)\label{eq:42}\end{eqnarray}
where $\widehat{A}\,=\widehat{A}^{\dagger}=ie_{l}\,\widehat{p}_{l}$
. We can now decouple equation (\ref{eq:42}) as 

\begin{eqnarray}
\widehat{A}\,\widehat{A}^{\dagger}\psi_{b}(x) & = & E^{2}\psi_{b}(x)\nonumber \\
\widehat{A}^{\dagger}\widehat{A}\,\psi_{a}(x) & = & E^{2}\psi_{a}(x)\label{eq:43}\end{eqnarray}
which gives rise to a single supersymmetric decoupled equation as

\begin{eqnarray}
P_{l}^{2}\psi_{a,b}(x) & = & E^{2}\psi_{a,b}(x)\label{eq:44}\end{eqnarray}
where $\psi_{a}(x)$ and $\psi_{b}(x)$ are eigen functions of partner
Hamiltonians $H_{-}=\widehat{A}^{\dagger}\widehat{A}$ and $H_{+}=\widehat{A}\,\widehat{A}^{\dagger}$.
The supersymmetric Hamiltonian is thus described as

\begin{eqnarray}
\widehat{H} & = & \left[\begin{array}{cc}
H_{+} & 0\\
0 & H_{-}\end{array}\right]=\left[\begin{array}{cc}
P_{^{l}}^{2} & 0\\
0 & P_{^{l}}^{2}\end{array}\right]=\hat{H_{D}^{2}}\label{eq:45}\end{eqnarray}
where $\hat{H_{D}}$ is the Dirac Hamiltonian given by 

\begin{eqnarray}
\hat{H_{D}} & =\sum_{l=1}^{3}\alpha_{l}p_{l}= & \left[\begin{array}{cc}
0 & ie_{l}p_{l}\\
ie_{l}p_{l} & 0\end{array}\right]\label{eq:46}\end{eqnarray}
and can be compared with the Dirac Hamiltonian given by Thaller \cite{key-27}
as 

\begin{eqnarray}
\hat{H_{D}} & = & \left[\begin{array}{cc}
M_{+} & Q_{D}^{\dagger}\\
Q_{D} & -M_{-}\end{array}\right]\label{eq:47}\end{eqnarray}
where we have obtained $M_{+}=M_{-}=0$ and $Q_{D}=Q_{D}^{\dagger}=i\, e_{l}p_{l}$
along with the following supersymmetric conditions

\begin{eqnarray}
Q_{D}^{\dagger} & M_{-}= & M_{+}Q_{D}^{\dagger}\,\,\,\,;\: Q_{D}M_{+}=M_{-}Q_{D}.\label{eq:48}\end{eqnarray}
Restricting the propagation along x-axis to discuss the quantum mechanics
in two dimensional space time, let us write 

\begin{eqnarray}
\widehat{p}_{l} & = & -i\,\frac{d}{dx}\nonumber \\
ie_{l}\,\widehat{p}_{l} & = & e_{2}\frac{d}{dx}\nonumber \\
\widehat{H} & = & \left[\begin{array}{cc}
-\frac{d^{2}}{dx^{2}} & 0\\
0 & -\frac{d^{2}}{dx^{2}}\end{array}\right]\label{eq:49}\end{eqnarray}
or\begin{eqnarray}
\widehat{H} & = & \left[\begin{array}{cc}
\widehat{Q}\widehat{Q}^{\dagger} & 0\\
0 & \widehat{Q}^{\dagger}\widehat{Q}\end{array}\right]\label{eq:50}\end{eqnarray}
where supercharges are described in terms of quaternion units i.e.

\begin{eqnarray}
\widehat{Q} & = & \left[\begin{array}{cc}
0 & -e_{2}^{\dagger}\frac{d}{dx}\\
0 & 0\end{array}\right]\nonumber \\
\widehat{Q}^{\dagger} & = & \left[\begin{array}{cc}
0 & 0\\
e_{2}\frac{d}{dx} & 0\end{array}\right]\label{eq:51}\end{eqnarray}
As such we may obtain the supersymmetry algebra as 

\begin{eqnarray}
\left[\widehat{Q\,},\widehat{H}\right] & = & \left[\widehat{Q\,},\widehat{H}^{\dagger}\right]=0\nonumber \\
\left\{ \widehat{Q\,},\widehat{Q\,}\right\}  & = & \left\{ \widehat{Q\,}^{\dagger},\widehat{Q\,}^{\dagger}\right\} =0\nonumber \\
\left\{ \widehat{Q\,},\widehat{Q\,}^{\dagger}\right\}  & = & \widehat{H}\label{eq:52}\end{eqnarray}
Here $\widehat{Q}^{\dagger},$ converts the upper component spinor
$\left\{ \begin{array}{c}
\psi_{a}\\
0\end{array}\right\} $to a lower one $\left\{ \begin{array}{c}
0\\
\psi_{b}\end{array}\right\} $and similarly $\widehat{Q}$ converts the lower component Spinor $\left\{ \begin{array}{c}
0\\
\psi_{b}\end{array}\right\} $to upper one $\left\{ \begin{array}{c}
\psi_{a}\\
0\end{array}\right\} $. If $\psi$ to be an eigen state of $H_{+}(H_{-})$, $\widehat{Q}$$\psi(\widehat{Q}^{\dagger}$$\psi)$
is the eigen state of that of $H_{+}(H_{-})$with equal energy.

\textbf{Case II}- \textbf{$m\,\neq0$ but constant and potential $\Phi=0$}.

Corresponding Dirac's equation (\ref{eq:35}) with its solution (\ref{eq:39})
for this case is described as

\begin{eqnarray}
(\gamma_{0}E\,-\gamma_{1}p_{1}-\gamma_{2}p_{2}-\gamma_{3}p_{3}-\, m) & \psi(x)= & 0\label{eq:53}\end{eqnarray}
Substituting $\gamma_{0}$ and $\gamma_{l}$ from equation (\ref{eq:36})
into equation (\ref{eq:53}) we get following coupled(supersymmetric)
equations

\begin{eqnarray}
ie_{l}p_{l}\psi_{a} & =\widehat{A}\psi_{a} & =(E+m)\psi_{b}\nonumber \\
ie_{l}p_{l}\psi_{b} & =\widehat{A}^{\dagger}\psi_{b} & =(E-m)\psi_{a}.\label{eq:54}\end{eqnarray}
which leads to following sets of decoupled equations,

\begin{eqnarray}
\widehat{A}^{\dagger}\psi_{b} & = & (E-m)\psi_{a}\nonumber \\
\widehat{A}\,\psi_{a} & = & (E+m)\psi_{b}\nonumber \\
\,\widehat{A}^{\dagger}\widehat{A}\,\psi_{a} & =H_{-}\psi_{a}=P_{l}^{2}\psi_{a}= & (E^{2}-m^{2}\,)\psi_{a}\nonumber \\
\widehat{A}\,\widehat{A}^{\dagger}\psi_{b} & =H_{-}\psi_{b}=P_{l}^{2}\psi_{b}= & (E^{2}-m^{2}\,)\psi_{b}\nonumber \\
\widehat{P_{l}}^{2}\psi_{a,b} & = & (E^{2}-m^{2}\,)\psi_{a,b}\label{eq:55}\end{eqnarray}
which are the Schrödinger equation for free particle.We may now write
the Schrödinger Hamiltonian and Schrödinger charges as 

\begin{eqnarray}
\widehat{H}_{s} & = & \left[\begin{array}{cc}
P_{l}^{2} & 0\\
0 & P_{l}^{2}\end{array}\right];\,\,\,\widehat{Q_{s}}=\left[\begin{array}{cc}
0 & ie_{l}p_{l}\\
0 & 0\end{array}\right];\,\,\,\widehat{Q_{s}}^{\dagger}=\left[\begin{array}{cc}
0 & 0\\
ie_{l}p_{l} & 0\end{array}\right].\label{eq:56}\end{eqnarray}
where $\widehat{H}_{s}$, $\widehat{Q_{s}}$ and $\widehat{Q_{s}}^{\dagger}$satisfy
the SUSY algebra given by equation (\ref{eq:52}). Correspondingly
the Dirac Hamiltonian is described as 

\begin{eqnarray}
\widehat{H}_{D}= & \sum_{l=1}^{3}\alpha_{l}p_{l}+m & =\left[\begin{array}{cc}
m & ie_{l}p_{l}\\
ie_{l}p_{l} & -m\end{array}\right]\label{eq:57}\end{eqnarray}
where $M_{+}=M_{-}=m$ and $\widehat{Q_{D}}=\,\widehat{Q_{D}}^{\dagger}=ie_{l}p_{l}$
and hence $\widehat{Q_{D}},\widehat{Q_{D}}^{\dagger},M_{+}$and $M_{-}$satisfy
the SUSY algebra given by equation (\ref{eq:52}). Thus the SUSY Hamiltonian
resembles with the square of Dirac Hamiltonian described as 

\begin{eqnarray}
\widehat{H} & = & \widehat{H}_{D}^{2}=\widehat{H}_{s}+m^{2}=\left[\begin{array}{cc}
\widehat{Q}\widehat{Q}^{\dagger} & 0\\
0 & \widehat{Q}^{\dagger}\widehat{Q}\end{array}\right]=\left[\begin{array}{cc}
\widehat{P_{l}}^{2}+m^{2} & 0\\
0 & \widehat{P_{l}}^{2}+m^{2}\end{array}\right]\label{eq:58}\end{eqnarray}
Similarly, we get the following relations while restricting ourselves
for two dimensional structure of space and time i.e.

\begin{eqnarray}
\widehat{H} & = & \left[\begin{array}{cc}
\widehat{Q}\widehat{Q}^{\dagger} & 0\\
0 & \widehat{Q}^{\dagger}\widehat{Q}\end{array}\right]=\left[\begin{array}{cc}
-\frac{d^{2}}{dx^{2}}+m^{2} & 0\\
0 & -\frac{d^{2}}{dx^{2}}+m^{2}\end{array}\right]\label{eq:59}\end{eqnarray}
where \begin{eqnarray}
\widehat{Q} & = & \left[\begin{array}{cc}
0 & -e_{2}^{\dagger}\frac{d}{dx}+m\\
0 & 0\end{array}\right]\nonumber \\
\widehat{Q}^{\dagger} & = & \left[\begin{array}{cc}
0 & 0\\
-e_{2}^{\dagger}\frac{d}{dx}+m & 0\end{array}\right]\label{eq:60}\end{eqnarray}
Hence we have restored the property of SUSY quantum mechanics and
obtain the commutation and anti commutation relations similar to that
of equation (\ref{eq:52}) for the free particle Dirac equation with
mass as well.

\textbf{Case III}- \textbf{Mass $m\neq0$ but not constant and is
space-time dependent i.e. $m=m(x)$ and} \textbf{potential $\Phi=0$-}

Let us discuss and verify the SUSY quantum mechanics for $m\neq0$
\textbf{i.e. $m=m(x)$} with scalar potential $\Phi=0$. We extend
the present theory in the same manner and express Dirac Hamiltonian
in the following form;

\begin{eqnarray}
\widehat{H_{D}} & = & \left[\begin{array}{cc}
0 & ie_{l}p_{l}-im\\
ie_{l}p_{l}-im & 0\end{array}\right]\label{eq:61}\end{eqnarray}

where $M_{+}=M_{-}=0,\widehat{Q_{D}}=ie_{l}p_{l}+im,\,\widehat{Q_{D}}^{\dagger}=ie_{l}p_{l}-im$
and again $\widehat{Q_{D}},\widehat{Q_{D}}^{\dagger},M_{+}$and $M_{-}$satisfy
the SUSY algebra given by equation (\ref{eq:52}).Thus operating Dirac
Hamiltonian given by equation (\ref{eq:61}) on the Dirac Spinor in
the manner $\widehat{H_{D}}\psi=E\psi$we get the following coupled
(supersymmetric) differential equations i.e.

\begin{eqnarray}
(ie_{l}p_{l}+im)\psi_{a} & = & \widehat{A}^{\dagger}\psi_{a}=E\psi_{b},\nonumber \\
(ie_{l}p_{l}-im)\psi_{b} & = & \widehat{A}\psi_{b}=E\psi_{a},\label{eq:62}\end{eqnarray}
where $\widehat{A}^{\dagger}=(ie_{l}p_{l}+im)$ and $\widehat{A}=(ie_{l}p_{l}-im)$
and hence we get the following form of Schrödinger Hamiltonian i.e.

\begin{eqnarray}
\widehat{H_{S}} & = & \left[\begin{array}{cc}
(ie_{l}p_{l}-im)(ie_{l}p_{l}+im) & 0\\
0 & (ie_{l}p_{l}-im)(ie_{l}p_{l}-im)\end{array}\right]=\widehat{H_{D}}^{2}\label{eq:63}\end{eqnarray}
which may be written as

\begin{eqnarray}
\widehat{H_{S}} & = & \left[\begin{array}{cc}
\widehat{Q_{s}}\widehat{Q}_{S}^{\dagger} & 0\\
0 & \widehat{Q_{S}}^{\dagger}\widehat{Q_{S}}\end{array}\right]=\widehat{H_{D}}^{2}\label{eq:64}\end{eqnarray}
with the following representation of super charges,

\begin{eqnarray}
\widehat{Q_{s}} & = & \left[\begin{array}{cc}
0 & (ie_{l}p_{l}-im)\\
0 & 0\end{array}\right]\nonumber \\
\widehat{Q_{S}}^{\dagger} & = & \left[\begin{array}{cc}
0 & 0\\
(ie_{l}p_{l}+im) & 0\end{array}\right].\label{eq:65}\end{eqnarray}
Equations(\eqref{eq:54} to\eqref{eq:65}) satisfy the supersymmetric
quantum mechanical relations given by equation (\ref{eq:47}) and
as such the supersymmetry.

\textbf{Case IV}- \textbf{Dirac equation in Electromagnetic Field}-
Before writing the quaternion Dirac equation in generalized electromagnetic
fields of dyons let us start with the quaternion gauge transformations.
A $Q-$field (\ref{eq:1}) is described in terms of following SO(4)
local gauge transformations\cite{key-6,key-9};

\begin{eqnarray}
\phi & \rightarrow\phi' & =U\,\phi\,\bar{V\,}\,\,\,\,\,\, U,V\,\varepsilon Q\,\,,\,\,\, U\,\bar{U\,}=V\,\overline{V}=1\label{eq:66}\end{eqnarray}
The covariant derivative for this is then written in terms of two
gauge potentials as 

\begin{eqnarray}
D_{\mu}\phi & = & \partial_{\mu}\phi+A_{\mu}\phi-\phi B_{\mu}\label{eq:67}\end{eqnarray}
where potential transforms as 

\begin{eqnarray*}
A_{\mu}^{'} & = & U\, A_{\mu}\bar{U}\,+U\,\partial_{\mu}\bar{U}\,\end{eqnarray*}
\begin{eqnarray}
B_{\mu}^{'} & = & V\, B_{\mu}\bar{V}\,+V\,\partial_{\mu}\bar{V}\,\label{eq:68}\end{eqnarray}
and

\begin{eqnarray}
\bar{\phi'}\phi' & =\overline{(U\phi\bar{V)}}((U\phi\bar{V)} & =\bar{\phi}\phi=\phi_{0}^{2}+\left|\vec{\phi}\right|^{^{2}}\label{eq:69}\end{eqnarray}
Here we identify the non Abelian gauge fields $A_{\mu}$and $B_{\mu}$as
the gauge potentials respectively for electric and magnetic charges
of dyons \cite{key-10,key-12,key-22,key-23}. Corresponding field
momentum of equation (\ref{eq:67}) may also be written as follows

\begin{eqnarray}
P_{\mu}\phi & = & p_{\mu}\phi+A_{\mu}\phi-\phi B_{\mu}\label{eq:70}\end{eqnarray}
where the gauge group $SO(4)=SU(2)_{e}\times SU(2)_{g}$ is constructed
in terms of quaternion units of electric and magnetic gauges .Accordingly,
the covariant derivative thus describes two different gauge field
strengths i.e.

\begin{eqnarray}
\left[D_{\mu},D_{\upsilon}\right]\phi & = & f_{\mu\nu}\phi-\phi h_{\mu\nu}\nonumber \\
f_{\mu\nu} & = & A_{\mu,\nu}-A_{\nu,\mu}+\left[A_{\mu},A_{\nu}\right]\nonumber \\
h_{\mu\nu} & = & B_{\mu,\nu}-B_{\nu,\mu}+\left[B_{\mu},B_{\nu}\right]\label{eq:71}\end{eqnarray}
where $f_{\mu\nu}$ and $h_{\mu\nu}$ are gauge field strengths associated
with electric and magnetic charges of dyons respectively.We may now
write the Dirac equation (\ref{eq:34}) as

\begin{eqnarray}
i\,\,\gamma_{\mu\,\,}D_{\mu}\psi(x,t) & = & m\,\psi(x,t).\label{eq:72}\end{eqnarray}
Following some restrictions and using the properties of quaternions
we may write the Dirac equation (\ref{eq:72}) as

\begin{equation}
\left[\begin{array}{cc}
m & ie_{l}(p_{l}+A_{l})\\
-ie_{l}(p_{l}+A_{l}) & -m\end{array}\right]\left[\begin{array}{c}
\psi_{a}\\
\psi_{b}\end{array}\right]+\left[\begin{array}{c}
\psi_{a}\\
\psi_{b}\end{array}\right]\left[\begin{array}{cc}
0 & -ie_{l}B_{l}\\
ie_{l}B_{l} & 0\end{array}\right]=(E-A_{0}+B_{0})\left[\begin{array}{c}
\psi_{a}\\
\psi_{b}\end{array}\right]\label{eq:73}\end{equation}
where $\varphi_{a}=\varphi_{0}+e_{1}\varphi$and $\varphi_{b}=\varphi_{2}-e_{1}\varphi_{3}.$
As such we obtain the following set of coupled equations responsible
for the supersymmetric realization of the theory i.e.\-\begin{eqnarray*}
E\psi_{a}= & (m+A_{0}-B_{0})\psi_{a}+ie_{l}(p_{l}+A_{l})\psi_{b}-i\psi_{b}e_{l}B_{l} & ;\\
E\psi_{b}= & (-m+A_{0}-B_{0})\psi_{b}-ie_{l}(p_{l}+A_{l})\psi_{a}+i\psi_{a}e_{l}B_{l} & ;\\
ie_{l}(p_{l}+A_{l})\psi_{b}-i\psi_{b}e_{l}B_{l} & = & (E-A_{0}+B_{0}-m)\psi_{a}\\
-ie_{l}(p_{l}+A_{l})\psi_{a}+i\psi_{a}e_{l}B_{l} & = & (E-A_{0}+B_{0}+m)\psi_{b}\end{eqnarray*}

\begin{eqnarray}
\widehat{A}^{\dagger}\psi_{b} & = & (E-m-A_{0}+B_{0})\psi_{a}=ie_{l}(p_{l}+A_{l})\psi_{b}-i\psi_{b}e_{l}B_{l}\nonumber \\
\widehat{A}\,\psi_{a} & = & (E+m-A_{0}+B_{0})\psi_{b}=-ie_{l}(p_{l}+A_{l})\psi_{a}+i\psi_{a}e_{l}B_{l}\nonumber \\
\widehat{A}^{\dagger}\widehat{A}\,\psi_{a} & = & (E^{2}-m^{2}\,)\psi_{a}\nonumber \\
\widehat{A}\,\widehat{A}^{\dagger}\psi_{b} & = & (E^{2}-m^{2}\,)\psi_{b}\label{eq:74}\end{eqnarray}
where we have restricted ourselves to the case of two dimensional
supersymmetry by imposing the condition $A_{1}^{\dagger}=-A_{1},A_{2}^{\dagger}=-A_{2},A_{3}^{\dagger}=-A_{3},B_{1}^{\dagger}=-B_{1},B_{2}^{\dagger}=-B_{2},B_{3}^{\dagger}=-B_{3}$
along with other subsidary conditions to restore the supersymmetry.
As such, it is possible to supersymmetrize the Dirac equation for
generalized electromagnetic fields of dyons and we may obtain the
commutation and anti commutation relations given by equation(\ref{eq:47})
to verify the supersymmetric quantum mechanics in this case.


\begin{thebibliography}{10}
\bibitem[1]{key-1} W. R. Hamilton, {}``Elements of quaternions'',
Chelsea Publications Co., NY, (\emph{1969}).

\bibitem[2]{key-2} L. Silberstein, Phil. Mag., \textbf{\underbar{63}}
(1912), 790. 

\bibitem[3]{key-3} P. G. Tait, {}``An elementary Treatise on Quaternions'',
Oxford Univ. Press. (1875).

\bibitem[4]{key-4} B. S. Rajput, S. R. Kumar and O. P. S. Negi, Letter.
Nuovo Cimento,\textbf{\underbar{34}} (1982), 180; \textbf{\underbar{36}}
(1983), 75.

\bibitem[5]{key-5}V. Majernik, PHYSICA, \textbf{\underbar{39}} (2000),
9 and references therein.

\bibitem[6]{key-6} S. L. Adler, {}``Quaternion Quantum Mechanics
and Quantum Fields'', Oxford Univ. Press, NY, (1995).

\bibitem[7]{key-7} D. Finklestein, J. M. Jauch, S. Schiminovich and
D. Speiser, J.Math. Phys., \textbf{\underbar{4}} (1963), 788.

\bibitem[8]{key-8} D. V. Duc and V. T. Cuong, Communication in Physics,
\textbf{\underbar{8}} (1998), 197.

\bibitem[9]{key-9} K. Morita, Prog. Theor. Phys., \textbf{\underbar{67}}
(1982), 1860; \textbf{\underbar{65}} (1981), 2071, 

\bibitem[10]{key-10} P. S. Bisht, O. P. S. Negi and B. S. Rajput,
Int. J. Theor. Phys., \textbf{\underbar{32}} (1993), 2099. 

\bibitem[11]{key-11} A. J. Davies, Phys. Rev., \textbf{\underbar{A49}}
(1994), 714. 

\bibitem[12]{key-12} Shalini Bisht, P. S. Bisht and O. P. S. Negi,
IL Nuovo Cimento, \textbf{\underbar{B113}}, (1998), 1449.

\bibitem[13]{key-13}A. Waser, AW-Verlag www.aw-Verlag.ch (2000);
V. V. Kravchenkov, {}``Applied Quaternion Analysis'', Helderman
Verlag, Germany (2003).

\bibitem[14]{key-14} M. F. Sohnius, Phys. Rep., \textbf{\underbar{128}}
(1985), 53.

\bibitem[15]{key-15} P. Fayet and S. Ferrara, Phys. Rep., \textbf{\underbar{32}}
(1977), 247.

\bibitem[16]{key-16} F. Cooper, A. Khare U. Sukhatme, Phys. Rep.,
\textbf{\underbar{251}} (1995), 267.

\bibitem[17]{key-17} V. A. Kostelecky and D. K. Campbell, {}``Supersymmetry
in Physics'', North Holland, (1985).

\bibitem[18]{key-18} A. Bilal, {}``Introduction to Supersymmetry'',
hep-th/0101055.

\bibitem[19]{key-19} E. Witten, Nucl . Phys., \textbf{\underbar{B202}}
(1982), 213.

\bibitem[20]{key-20} C. M. Hull, {}``The geometry of Supersymmetric
Quantum Mechanics'', hep-th/9910028.

\bibitem[21]{key-21} A. Das,S. Okubo and S.A. Pernice, Mod. Phys.
Lett., \textbf{\underbar{A12}} (1997) 581; {}``Higher Dimensional
SUSY Quantum Mechanics'', hep-th/9612125. 

\bibitem[22]{key-22} B. S. Rajput and D. C. Joshi, Had. J., \textbf{\underbar{4}}
(1981), 1805.

\bibitem[23]{key-23} B. S. Rajput and Om Prakash, Indian J. Phys.,
\textbf{\underbar{A53}} (1979), 274.

\bibitem[24]{key-24} P. Roteli, Mod.Phys. Lett., \textbf{\underbar{A4}}
(1989)933; A4(1989) 1763.

\bibitem[25]{key-25} A. J. Davies, Phys. Rev., \textbf{\underbar{D41}}(1990),2628;
A.Govorkov,Theor.Math.Phys., \textbf{\underbar{68}} (1987) 893.

\bibitem[26]{key-26} H. Osborn, Phys. Lett., \textbf{\underbar{B83}}
(1979) 321; P.de Vecchia, {}``Duality in Supersymmetric Gauge Theories'',
hep-th/9608090 and references therein.

\bibitem[27]{key-27} B. Thaller, {}``The Dirac Equation'', Springer
verlag, Berlin (1992); {}``The Dirac particle in magnetic field'',
in A. Boutet de Monvel, P. Dita, G. Nenciu, and R. Purice Eds., 'Recent
Developments in Quantum Mechanics', Mathemetical Physics Studies Nr.
\textbf{\underbar{12}} (Kluwer Acad. Publ. Dodrecht,1991) p.351-366.
\end{thebibliography}
\end{document}